\documentclass[aps,showpacs,prd,twocolumn,superscriptaddress,floatfix]{revtex4}
\usepackage{amsmath, amsfonts, hyperref}
\input epsf
\usepackage{graphicx}
\newcommand{\beq}{\begin{equation}}
\newcommand{\eeq}{\end{equation}}
\newcommand{\beqn}{\begin{eqnarray}}
\newcommand{\eeqn}{\end{eqnarray}}

\newcommand{\apjl}{Astrophys.\ J.\ Lett.}
\newcommand{\mnras}{Mon.\ Not.\ R.\ Astro.\ Soc.}

\newcommand{\apjs}{Astrophys.\ J.\ Supp.}

\begin{document}
\title
{Black holes, disks and jets following binary mergers and stellar collapse: \\ 
The narrow range of EM luminosities and accretion rates}
\date{\today}
\author{Stuart~L. Shapiro}
\altaffiliation{Also Department of Astronomy and NCSA, University of
  Illinois at Urbana-Champaign, Urbana, IL 61801}
\affiliation{Department of Physics, University of Illinois at Urbana-Champaign, Urbana, IL 61801}

\begin{abstract}

We have performed magnetohydrodynamic simulations in general relativity of
binary neutron star and binary black hole-neutron star mergers, as well
as the magnetorotational collapse of supermassive stars. In many cases 
the outcome is a spinnng black hole (BH) immersed in a magnetized disk, 
with a jet emanating from the poles of the BH.  
While their formation scenarios differ and their BH masses, as well as their 
disk masses, densities, and magnetic field strengths, vary by 
orders of magnitude, these features conspire to
generate jet Poynting luminosities that all lie in the same, narrow range of
$\sim 10^{52\pm1}~{\rm erg ~s^{-1}}$. A similar result applies to their
BH accretion rates upon jet launch, which is 
$\sim 0.1-10~{\rm M_{\odot}~s^{-1}}$. 
We provide a simple model that explains these unanticipated findings.
Interestingly, these luminosities reside in the same narrow range characterizing
the observed luminosity distributions of over 400 short and long GRBs
with distances inferred from spectroscopic redshifts or host galaxies.
This result, together with the GRB lifetimes predicted by the model, 
supports the belief that a compact binary
merger is the progenitor of an SGRB, while a massive, stellar 
magnetorotational collapse is the progenitor of an LGRB.

\end{abstract}

\pacs{04.25.D-, 04.25dg, 98.70.Rz, 47.75.+f}
\maketitle

\section{Introduction and Motivation} 

We have recently performed simulations of 
merging binary neutron stars (NSNSs)~\cite{RLPS16} and 
binary black hole-neutron stars (BHNSs)~\cite{PRS15} 
in full general relativistic magnetohydrodynamics 
(GRMHD). The neutron stars (NSs) were threaded by dynamically weak 
dipole magnetic fields initially.
These simulations revealed the launching
of jets from the poles of the BH-disk remnants
following the mergers. 
Our simulations began from the late binary 
inspiral phase, continued through the NS tidal disruption and merger phases and 
did not terminate until the magnetized disk of NS debris reached a 
state of quasistationary accretion onto the remnant BH.
These are the first GRMHD simulations 
that demonstrated the launching of {\it bonafide} incipient jets 
following such mergers, i.e., outward streams of plasma
from the poles of the spinning BH remnants, with flows 
confined and driven outward in a narrow beam by a 
collimated, tightly wound, helical magnetic field
(see \cite{RLPS16,PRS15,P16} for summaries of and references to this and 
earlier work). The plasma jet is accompanied by
beam of electromagnetic (EM) Poynting radiation. 
This demonstration lends support to the sugggestion~\cite{ELPS89,NPP92,MHIM93} 
and widely-held notion that these mergers are the 
engines that power short gamma-ray bursts (SGRBs).  If true, then 
observable EM radiation typical of SGRBs could accompany the 
gravitational waves (GWs) detected from such events.

We also have performed GRMHD simulations of the
magnetorotational collapse of a supermassive star (SMS)~\cite{SPRS17}
triggered by a relativistic dynamical instability at the endpoint of
stellar evolution~\cite{Cha64a,Cha64b,Fey64,BS99}. 
Such stars could provide the seeds of the 
supermassive black holes (SMBHs) that reside in most galaxies, 
including the Milky Way, and are believed to power 
AGNs and quasars~\cite{Ree84}. 
These simulations, which can be scaled to stars of arbitrary mass, 
may also furnish 
crude models of the collapse of very 
massive Pop III stars, alternative sources of SMBH 
seeds~\cite{MadR01}. Scaling the mass downwards to  ``collapsars'', they
may provide simplified models of long GRBs (LGRBS)~\cite{MacW99}.
We shall collectively refer to these simulations as the SMS scenario.
The simulations, performed in full 3+1 dimensions, 
showed that, upon arriving at the onset of instability, 
an SMS threaded by a dynamically weak magnetic field and spinning uniformly 
at the mass-shedding limit collapses to a 
spinning BH remnant immersed in a magnetized accretion disk.
The results found for the BH and disk masses and BH spins were in 
good agreement with all previous GR simulations 
that started with the same uniformly rotating, unstable star and
adopted the same radiation pressure-dominated EOS. But all such previous 
simulations were performed in axisymmetry 
(see, e.g., ~\cite{ShiS02,ShiSUU16,LiuSS07}), and
only~\cite{LiuSS07} incorporated magnetic fields (stellar interior only).
These latest 3D simulations also agreed with analytic 
predictions~\cite{ShaS02} for the key remnant parameters.
However, by threading the star with a dynamically weak 
dipolar magnetic field and evolving it for many dynamical timescales
following collapse ($\Delta t \gtrsim 30,000M$) 
these 3D simulations launch a jet from the poles of the BH once 
the disk has settled down to a state of quasistationary accretion. Once again, 
the GW burst is accompanied by an appreciable EM Poynting flux.

In both our compact binary merger and SMS collapse scenarios we showed that
the EM power generated by the jets is likely the result of the 
Blandford-Znajeck (BZ) mechanism~\cite{BlaZ77}. In such a case the
Poynting luminosity is given roughly by
\begin{equation}
\label{LBZ}
L_{\rm BZ} \sim B_{\rm p}^2 M_{\rm BH}^2 
      \left( \frac{a}{M_{\rm BH}} \right)^2 
      \Big[ {\cal L}_0 \Big]
       \sim 10^{52 \pm 1}~{\rm erg~s^{-1}}
\end{equation}
where $B_{\rm p}$ is the poloidal $B$-field above the BH poles, $M_{\rm BH}$ is the BH mass,
and $a/M_{\rm BH}$ is the BH spin parameter, all in gravitational
units with $G=c=1$. The corresponding quasistationary rest-mass 
accretion rate is given roughly by
\begin{equation}
\label{Mdot}
\dot{M}_{\rm BH} \sim 4 \pi \rho M_{\rm BH}^2
       \Big[ {\dot{\cal M}_0} \Big] 
        \sim 0.1-10~M_{\odot}~{\rm s^{-1}},
\end{equation} 
where $\rho$ is the rest-mass density near the BH horizon. The quantities
${\cal L}_0$ and $\dot{\cal{M}}_0$ appearing in Eqs.~(\ref{LBZ}) and (\ref{Mdot})
restore the factors of $G$ and $c$ and with them the physical dimensions for $L_{\rm BZ}$ and 
$\dot{M}_{\rm BH}$:
\begin{eqnarray}
{\cal L}_0 &\equiv& c^5/G = 3.6 \times 10^{59} ~{\rm erg~s^{-1}}, \\
 {\dot{\cal M}_0} &\equiv& c^3/G = 2.0 \times 10^5 ~M_{\odot}~{\rm s^{-1}}. \nonumber
\end{eqnarray}
 
The collective numerical results of our simulations pose the following puzzle:
{\it why do the physical magnitudes of $L_{\rm BZ}$ and $\dot{M}_{\rm BH}$ 
found for the different remnant BH-disk-jet systems in the 
NSNS, BHNS and SMS cases we have simulated all turn out to be comparable?}
The fact is that their physical values are {\it all} within the 
rather narrow ranges 
indicated on the right-hand sides of Eqs.~(\ref{LBZ}) and (\ref{Mdot}). 
This finding is most unexpected, as the scenarios leading the formation of the
BH-disk-jet systems are very different and, more significantly, the masses, 
length and time scales, and the $B$-fields and densities characterizing 
these systems can differ by {\it many} 
orders of magnitude, particularly when we consider SMS progenitors. 
For example, the compact binary mergers form 
stellar mass BHs, while the SMS collapse
scenario applies to the formation of arbitrarily large SMBH masses. How do we explain why
their values of $L_{\rm BZ}$ and $\dot{M}_{\rm BH}$ are all within an order of 
magnitude or two of each other in light of the middle expressions in 
Eqs.~(\ref{LBZ}) and (\ref{Mdot}), which  exhibit quadratic scaling with
$M_{\rm BH}$, as well as quadratic scaling with $B_p$ and linear
scaling with $\rho$? The remnant SMBH masses can vary over many decades for
different SMS mass progenitors, and they can differ by orders of magnitude
from the masses in the compact binary scenarios. Similar differences 
apply to $B_p$ and $\rho$.  How can these parameters conspire to generate 
similar luminosities and similar rates of accretion for the
initial data we evolved?

We will resolve this puzzle below and discuss 
some of its implications. Unless explicitly noted
otherwise, we adopt geometrized
units with $G=1=c$.

\section {Explanation}

Recall that the BZ luminosity measures a Poynting flux
$\vec{S} = (\vec{E} \times \vec{B})/4 \pi$
propagating out from above the BH poles and is given, very
roughly,  by
\begin{equation}
\label{LBZa}
L_{\rm BZ} \sim S \pi r_{\rm H}^2 \sim \frac{B^2_{\rm p}}{4 \pi} \pi 
        r_{\rm H}^2 \sim B^2_{\rm p} M_{\rm BH}^2, 
\end{equation}
where $r_{\rm H}$ is the BH horizon radius.
Since no flux is generated for nonspinning BHs, 
there must also be a  dependence on the BH spin 
parameter. So more precisely, 
combining  $\vec{S}$ with
the MHD relation $\vec{E}=-\vec{v} \times \vec{B}$ for a rotating
cylindrical region of gas above a BH pole gives a Poynting luminosity
$L_{\rm Poyn} \sim (B_{\rm p} B_{\phi}/4 \pi) (r_{\rm H} \Omega_{\rm H})
          (\pi r_{\rm H}^2)$
Here $\Omega_{\rm H} = a/(2M_{\rm BH} r_{\rm H})$
is the angular velocity of the horizon.
The maximum rate at which this flux can be generated occurs for the
maximal value of $B_{\phi}$,
$B_{\phi}(\rm max) \sim (r_{\rm H} \Omega_{\rm H})B_{\rm p}$
~(see, e.g., \cite{LivOP99}, Eq.2)
which gives the BZ luminosity quoted in
~\cite{ThoPM86}, Eq. (4.50),
\begin{equation}
\label{LBZexact}
L_{\rm BZ} \sim 
\frac{1}{128} B_p^2 r_{\rm H}^2 
\left( \frac{a}{M_{\rm BH}} \right)^2,
\end{equation}
or, scaled to the BHNS remnant parameters in the simulations of~\cite{PRS15},
\begin{equation}
\label{LBZcgs}
L_{\rm BZ}  \sim 10^{51} 
        \left( \frac{B_{\rm p}}{10^{15}~{\rm G}} \right)^2
        \left( \frac{M_{\rm BH}}{5.6~M_{\odot}} \right)^2
        \left( \frac{a}{M_{\rm BH}} \right)^2~{\rm erg \ s^{-1}}.
\end{equation}
 
Now an outward jet can form only when the magnetic field is
sufficiently strong to confine the plasma, reverse the inflow above the 
BH poles and drive an outflow. This ability requires the 
polar field  to be force-free,
$B^2_{\rm p}/8 \pi \rho_{\rm p} \gtrsim 1$, which when satisfied
is found to trigger the launching of a jet within a tightly wound, collimated, 
helical magnetic funnel above the poles. The force-free condition is 
satisfied once the plasma density above the poles
has been sufficiently reduced and the initial magnetic field
has been sufficiently amplified. Magnetic winding, 
MRI and the Kelvin-Helmholtz instability all 
contribute~\cite{PRS15,KiuCKSS15,RLPS16,SPRS17}. 
The magnetic field exhibits a coherent polar 
component with a net vertical flux threading the BH horizon.
The characteristc magnitude of the plasma density is 
established very roughly by the mean density in the disk,
\begin{eqnarray}
\label{rhoexact}
\rho &\sim& \frac{M_{\rm disk}}{2 H_{\rm disk} \pi R_{\rm disk}^2} \\
&\sim& \left( \frac{1}{2\pi} \right) 
\left( \frac{M_{\rm disk}}{M_{\rm BH}} \right)
\left( \frac{M_{\rm BH}}{R_{\rm disk}} \right)^2
\left( \frac{M_{\rm BH}}{H_{\rm disk}} \right) M_{\rm BH}^{-2}. \nonumber
\end{eqnarray}
The disk is a bloated toroid of half-thickness $H_{\rm disk}$ and outer 
radius $R_{\rm disk}$ that engulfs the BH. The radius is determined by the
angular momentum of the gas, which is set by the 
outermost layers of the NS following tidal break-up for binary mergers and by 
the angular momentum in the spinning outer envelope of the SMS progenitor 
for magnetorotational collapse. In all cases the disks are geometrically 
thick. For typical cases, whenever a jet is launched, $\rho_{\rm p}$ 
has fallen to values less than $\sim 0.1- 0.01$ times the mean density $\rho$ 
in the disk.  At the same time the ratio $B^2_{\rm p}/8 \pi \rho_{\rm p}$ 
grows to over $10-100$ in the jet once it is fully established. 
So very crudely these two factors cancel and we can write
\begin{equation}
\label{B}
B^2_{\rm p} \sim 8 \pi \rho
\end{equation}
to relate $B_{\rm p}$ near the BH poles to $\rho$ in the disk. 
In this spirit
we will neglect the detailed toroidal geometry of the disk and simply set
$2 H_{\rm disk} \sim R_{\rm disk}$ in Eq.~(\ref{rhoexact}) to get
\begin{equation}
\label{rho}
\rho \sim 
\frac{M_{\rm disk}}{\pi R_{\rm disk}^3} \sim 
\frac{1}{\pi}
\left( \frac{M_{\rm disk}}{M_{\rm BH}} \right)
\left( \frac{M_{\rm BH}}{R_{\rm disk}} \right)^3 
M_{\rm BH}^{-2}.
\end{equation}
Combining Eqs.~(\ref{B}) and (\ref{rho}) yields
\begin{equation}
\label{BM}
B_p^2 M_{\rm BH}^2 \sim 8 \pi \rho M_{\rm BH}^2 
\sim 8 \left( \frac{M_{\rm disk}}{M_{\rm BH}} \right)
\left( \frac{M_{\rm BH}}{R_{\rm disk}} \right)^3. 
\end{equation}
Inserting Eq.~(\ref{BM}) into Eq.~(\ref{LBZexact}) now yields
\begin{equation}
\label{LBZb}
L_{\rm BZ} \sim 
\frac{1}{10}
\left( \frac{M_{\rm disk}}{M_{\rm BH}} \right)
\left( \frac{M_{\rm BH}}{R_{\rm disk}} \right)^3
\left( \frac{a}{M_{\rm BH}} \right)^2
\Big[ {\cal L}_0 \Big],
\end{equation}
where employing gravitational units makes each factor in parenthesis nondimensional, 
so that inserting ${\cal L}_0$ again 
restores physical dimensions to $L_{\rm BZ}$.
Similarly inserting Eq.~(\ref{BM}) into Eq.~(\ref{Mdot}) gives
\begin{equation}
\label{Mdota}
\dot{M}_{\rm BH} \sim 4 
\left( \frac{M_{\rm disk}}{M_{\rm BH}} \right)
\left( \frac{M_{\rm BH}}{R_{\rm disk}} \right)^3
\Big[ {\dot{\cal M}_0} \Big] 
\end{equation} 

Eq.~(\ref{BM}) provides the crucial relations: this equation reveals that the products  
$B_p^2 M_{BH}^2$ and $\rho M_{BH}^2$ appearing in Eqs.~(\ref{LBZ}) and (\ref{Mdot}) are 
comparable in magnitude and determined by 
the disk mass and size normalized to the 
corresponding remnant BH parameters.  
For all the different compact binary merger and SMS collapse simulations that
we have found to launch {\it bonafide} incipient jets, these nondimensional, {\it normalized} disk mass and size ratios fall 
within a narrow range. For this reason it is no longer so 
surprising why the values of ${\cal L}_{\rm BZ}$ and $\dot{M}_{\rm BH}$, 
given by Eqs.~(\ref{LBZb}) and (\ref{Mdota}), also fall 
in a narrow range. We give examples in the next section.

Combining Eqs.~(\ref{LBZb}) and (\ref{Mdota}) yields the 
Poynting radiation efficiency $\epsilon$:
\begin{equation}
\label{eps}
\epsilon = \frac{L_{\rm BZ}}{\dot{M}_{\rm BH}} \sim \frac{1}{40}  
\left( \frac{a}{M_{\rm BH}} \right)^2.  
\end{equation}
The Poynting efficiency for the jet is less than 
the EM efficiency for a corotating, {\it thin} Keplerian disk accreting 
onto a rapidly spinning BH and is zero for a nonspinning Schwarzschild BH.
However, the EM emission from the {\it thick}   
toroids found here may be less efficient and will likely appear with a  
different spectrum, particularly if  
the Poynting flux from the jets is ultimately pumped into prompt $\gamma$-rays.

\section {Applications}
\subsection{NSNS mergers}

Consider the merger of identical, magnetized, irrotational NSNSs in circular orbit and obeying a stiff $\Gamma-$law EOS with $\Gamma=2$. Take each star to have a fraction 0.81 of the rest-mass
of the maximum-mass TOV star constructed from the same EOS and thread them 
with a dynamically weak, interior, dipole $B$-field 
(i.e, pressure ratio $P_{\rm mag}/P_{\rm gas} \ll 1$) and a strong, 
exterior dipole field (i.e., a pulsarlike magnetosphere, 
$P_{\rm mag}/P_{\rm gas} \gg 1$)~\cite{RLPS16}. The merger results in the 
formation of a hypermassive neutron star (HMNS) 
that undergoes delayed collapse to a spinning
black hole immersed in a magnetized disk. The delayed collapse of the
HMNS is crucial for the buildup of the magnetic field and jet 
launch, which do not occur for prompt 
collapse characterizing more massive 
NSNSs~\cite{RPS17}. We extract 
the following results from the simulation 
for the nondimensional ratios appearing in Eqs.~(\ref{BM})-(\ref{Mdota})
after jet launch:
\begin{equation}
\label{NSNSdata}
\frac{M_{\rm disk}}{M_{\rm BH}} \sim 1.1 \times 10^{-2}, \ \ \frac{R_{\rm disk}}{M_{\rm H}} \sim 56, 
\ \ \frac{a}{M_{\rm BH}} \sim 0.74 .
\end {equation}
Most of the rest-mass goes into the remnant BH following collapse of the HMNS. 
Inserting the above ratios into Eqs.~(\ref{BM})-(\ref{Mdota}) yields the 
model values listed in the first row of Table~\ref{tab:NSNS}. 
The second row displays the values extracted
from the simulation data following jet launch. Given the approximate nature
of the above model equations, where the numerical coefficients are crude estimates
at best, and given the raw extraction of time-varying simulation data from representative 
snapshots, we quote only the orders of magnitude of the tabulated quantities.
Allowing for the rough nature of our simple analysis,  
the model expectations are consistent with the simulation data.

\begin{table}[t]
\caption{\label{tab:NSNS} NSNS remnant after jet launch} 
    \centering
\begin{tabular}{ccccc}
\hline\hline
Source  & $\rho \ {\rm (g/cm^3)}$
\footnote{$\rho$ and $B_{\rm p}^2$ scale
as $(M_{\rm BH}/2.85 M_{\odot})^{-2}$}   
& $B_{\rm p} \ {\rm (G)~^a}$ & $\dot{M}_{\rm BH} \ {\rm (M_{\odot}/s)}$ 
& $L_{\rm BZ}$ (erg/s)\\
\hline
model\footnote{Eqs.~(\ref{BM})-(\ref{Mdota},\ref{NSNSdata})}   
&    $10^9$        &  $10^{15}$   & 0.1    & $10^{51}$  \\
simulation\footnote{\cite{RLPS16}}    
&  $10^9$          &  $10^{16}$   &  0.1     &  $10^{51}$   \\
\hline\hline
\end{tabular}
\end{table}

\subsection{BHNS mergers}

Now consider the merger of a magnetized, irrotational NS in 
circular orbit about a 
spinning BH three times its mass~\cite{PRS15}. The NS, which again obeys
a $\Gamma=2$ EOS, is a fraction 0.83 of the rest-mass
of the maximum-mass TOV star and is again threaded 
by a dynamically weak, interior, dipole $B$-field and a strong, 
pulsarlike, exterior dipole field.
The black hole has an initial spin $a/M_{\rm BH} = 0.75$ aligned with orbital
angular momentum.
We extract the following results from the simulation
for the nondimensional ratios appearing in Eqs.~(\ref{BM})-(\ref{Mdota})
after jet launch:
\begin{equation}
\label{BHNSdata}
\frac{M_{\rm disk}}{M_{\rm BH}} \sim 2.5 \times 10^{-2}, \ \ \frac{R_{\rm disk}}{M_{\rm BH}} \sim 30, 
\ \ \frac{a}{M_{\rm BH}} \sim 0.85 .
\end {equation}

When we insert the above ratios into Eqs.~(\ref{BM})-(\ref{Mdota} and obtain
the values recorded in Table~\ref{tab:BHNS} we find that the model is again consistent with 
the simulation data.

\begin{table}[t]
\caption{\label{tab:BHNS} BHNS remnant after jet launch}
    \centering
\begin{tabular}{ccccc}
\hline\hline
Source  & $\rho \ {\rm (g/cm^3)}$
\footnote{$\rho$ and $B^2_{\rm p}$ scale as $(M_{\rm BH}/5.6 M_{\odot})^{-2}$}
& $B_{\rm p} \ {\rm (G)^a}$ & $\dot{M}_{\rm BH} \ {\rm (M_{\odot}/s)}$
& $L_{\rm BZ}$ (erg/s)\\
\hline
model\footnote{Eqs.~(\ref{BM})-(\ref{Mdota},\ref{BHNSdata})}
&    $10^{10}$        &  $10^{16}$   & 1    & $10^{52}$  \\
simulation\footnote{\cite{PRS15}}
&  $10^{10}$          &  $10^{15}$   &  0.1     &  $10^{51}$   \\
\hline\hline
\end{tabular}
\end{table}

\subsection{SMS collapse}

\begin{table}[t]
\caption{\label{tab:SMS} SMS remnant after jet launch}
    \centering
\begin{tabular}{ccccc}
\hline\hline
Source  & $\rho \ {\rm (g/cm^3)}$
\footnote{$\rho$ and $B^2_{\rm p}$ scale as $(M_{\rm BH}/10^6 M_{\odot})^{-2}$}
& $B_{\rm p} \ {\rm (G)^a}$ & $\dot{M}_{\rm BH} \ {\rm (M_{\odot}/s)}$
& $L_{\rm BZ}$ (erg/s)\\
\hline
model\footnote{Eqs.~(\ref{BM})-(\ref{Mdota}),\ref{SMSdata})}
&    $10^{-2}$        &  $10^{10}$   & 0.1    & $10^{51}$  \\
simulation\footnote{\cite{SPRS17}}
&  $0.1$          &  $10^{11}$   &  1     &  $10^{52}$   \\
\hline\hline
\end{tabular}
\end{table}

Consider finally the magnetorotational collapse of an SMS 
rotating uniformly at the mass-shedding limit and marginally unstable to a 
general relativistic dynamical (radial) instability~\cite{SPRS17}. The
star is governed by a $\Gamma$-law, radiation pressure-dominated EOS 
with  $\Gamma=4/3$ and is again threaded
by a dynamically weak, interior, dipolar $B$-field and a 
strong, pulsarlike, exterior field,
all aligned with the spin axis. The mass of the star is 
arbitrary. About 92\% of the mass goes into
forming a spinning black hole, with most of 
the rest comprising a bloated, toroidal disk. 
Eventually jets of gas confined by collimated, helical magnetic 
fields emerge from both poles of the BH. A simulation yields the
following nondimensional ratios after jet launch:
\begin{equation}
\label{SMSdata}
\frac{M_{\rm disk}}{M_{\rm BH}} \sim 0.1, \ \ \frac{R_{\rm disk}}{M_{\rm BH}} \sim 100, 
\ \ \frac{a}{M_{\rm BH}} \sim 0.68 .
\end {equation}

Using the data in Eq.~(\ref{SMSdata}), 
Eqs.~(\ref{BM})-(\ref{Mdota}) yield
the results recorded in Table~\ref{tab:SMS}, which are 
again compared with the data extracted from our SMS simulation. 
Once again the model expectations are roughly consistent with the
numerical results. This SMS scenario for BH-disk-jet formation
affords a strong test of our model, as the mass, length, timescales and 
magnetic field strengths all differ by orders of magnitude from
those characterizing the compact binary merger scenarios. A comparison of the values tabulated for $\rho$ and $B$ bear this out. 
In fact, these parameters can vary by orders of magnitude even within the SMS 
scenario itself by allowing
the SMS progenitor mass to vary by orders of magnitude.  Yet the 
predicted values of  $L_{\rm BZ}$ and $\dot{M}_{\rm BH}$
for the compact binary merger and SMS collapse scenarios
are within an order of magnitude of each other. 
This establishes the near-universality of these quantities, at least for the 
BH-disk-jet formation mechanisms we have simulated. For our SMS 
collapse scenario alone,  $L_{\rm BZ}$ and $\dot{M}_{\rm BH}$ 
are {\it strictly} independent of the progenitor mass, assuming a
fixed initial magnetic field topology and ratio of magnetic to gravitational
potential energy.
 
\section {Implications}

The restriction of jet Poynting luminosities and disk 
rest-mass accretion rates to 
a narrow range in our simulations was unanticipated, 
but is no longer a mystery. 
Even within our model we still expect some variation in these 
quantities for different merger or collapse cases,
as the disk and black hole spin ratios appearing on 
right-hand sides of Eqs.~(\ref{LBZb}) 
and (\ref{Mdota}) will never be identical for different scenarios leading
to BH-disk-jet systems. 
But each of these ratios can never exceed unity, so that our model predicts
that the extreme upper limit, never approached in our simulations to date, 
is $\sim {\cal L}_0$ for
the Poynting luminosity and $\sim {\cal \dot M}_0$ for the accretion rate.
The fact that the luminosities and accretion rates in our simulations 
are universally 6-7 orders of magnitude 
smaller simply reflects the fact that in all cases
the disk mass is typically $\lesssim 0.1$
times smaller than the BH mass and the disk radius is larger 
than the BH horizon radius by a factor $\gtrsim 50$. 

There are several distinguishing, observable features 
that are expected to be quite different for the different cases. 
One of these is the disk
lifetime, which we can estimate using Eq.~(\ref{Mdota}):
\begin{equation}
\label{tdisk} 
t_{\rm disk}/M_{\rm BH} \sim \frac{(M_{\rm disk}/M_{\rm BH})}{\dot{M}_{\rm BH}} 
          \sim \frac{1}{4} \left( \frac{R_{\rm disk}}{M_{\rm BH}} \right)^3
          \Big[ \frac{1}{ {\dot{\cal M}_0}} \Big]. 
\end{equation}
Substituting Eqs.~(\ref{NSNSdata})-(\ref{SMSdata}) into Eq.~(\ref{tdisk}) then gives 
\begin{eqnarray}
\label{tdiskb}
t_{\rm disk} &\sim& 0.6~(M_{\rm BH}/2.85 M_{\odot})~{\rm s}  \ \ \ \ \ \ \  {(\rm NSNS}) \\ 
             &\sim& 0.2~(M_{\rm BH}/5.6 M_{\odot})~{\rm s}     \ \ \ \ \ \ \ \ \ {(\rm BHNS}) \nonumber \\ 
             &\sim& 1 \times 10^6~(M_{\rm BH}/10^6 M_{\odot})~{\rm s}  \ \ \ {(\rm SMS)}. \nonumber
\end{eqnarray}

If the BH-disk-jet remnants in these systems are the engines that power 
GRBs, then the disk lifetimes computed above (and not simply the compact binary 
merger or stellar collapse timescales) might be directly related to
the observed prompt $\gamma$-ray burst (rest-frame) lifetimes.
Eqn.~(\ref{tdiskb}) is then consistent with the widely-held view that 
NSNS and/or BHNS mergers power SGRBs, with burst timescales $\lesssim 2$~s.
The widely-held view  that
magnetorotational, massive stellar collapse may give rise to LGRBs,
with burst timescales $\gtrsim 2$~s,
is also consistent with Eq.~(\ref{tdiskb}), provided our crude 
magnetorotational  collapse model can be extended downward in mass to 
stars that collapse to black holes with 
$M_{\rm BH}  \gtrsim 10~M_{\odot}$ (``collapsars"~\cite{MacW99}).  
Our model suggests that such collapse 
will give rise to Poynting luminosities very 
comparable to those for compact binary mergers, but 
with longer burst timescales  
because $M_{\rm BH}$ and $(R_{\rm disk}/M_{\rm BH})^3$ are both 
larger. As the disk is consumed and the BH accretion rate falls,
the jet luminosity presumably will decline rapidly, as in observed GRBs and in
some MAD accretion models (see, e.g., ~\cite{TchG15} and references therein).

A recent survey~\cite{LiZL16} of 407 GRBs, which includes 375 GRBs
with spectroscopic redshift measurements and 32 others with host
galaxy information, is most revealing. 
Table 1 and the panels in Fig.~1 of this survey show
a fairly narrow width of a couple of orders of magnitude 
centered at $\gtrsim 10^{52}$~erg/s in the distributions of the isotropic 
peak (prompt) $\gamma$-ray luminosities of both SGRBS and LGRBs 
The narrow range of luminosities in our simulations 
centered near this value and explained with our simple model is 
consistent with this data. (The agreement might be improved further 
by correcting for beaming.) The magnitude of this luminosity 
enables these sources to be observed out to high 
redshift ($z_{\rm max}  = 8.23$) by present
$\gamma$-ray satellites. This property bodes well for future instruments
designed to use GRBs at high redshift as a tool to probe the early 
Universe~\cite{Sal15}.

Simulations describing alternative scenarios for SMS collapse also have
been performed in GR. Consider a 
rapidly {\it differentially} rotating star whose collapse is triggered
by lowering $\Gamma$ to 1.33 to account for $e^+e^-$ pair 
production~\cite{Rei13}.
When the star is seeded by a small initial $m=2$ density perturbation
it forms a binary SMBH during the collapse. The binary ultimately
merges, forming a single BH-disk system.
No magnetic fields are incorporated, so there is no jet. Had the star
been threaded by a dynamically
weak magnetic field and a jet formed, the expected
Poynting luminosity can be inferred from Eqn.~(\ref{eps}), using 
the reported accretion rate (${\dot M} = 6.7 \times 10^{-5} {\dot{\cal M}_0} 
= 13~M_{\odot}$/s) 
and remnant BH spin ($a/M_{\rm BH} = 0.9$): $L_{\rm BZ} \sim 10^{53}$ erg/s. 
This estimate falls in the narrow range of Poynting luminosities 
predicted by our model for BH-disk-jets formed 
following stellar collapse. It suggests that  
the GW signals calculated in \cite{Rei13} could be accompanied
by counterpart EM bursts, were a magnetic field present. 
Detailed simulations in GRMHD are required to confirm these expectations.

While the {\it Fermi} detection~\cite{Con16} of a weak SGRB during
the LIGO binary black hole (BHBH) event GW150914~\cite{Abb16} may
be a chance coincidence, there are scenarios whereby a BHBH merger could
occur within an ambient disk. If any scenario results
in a BH-disk-jet remnant similar to those found in our simulations, 
then our model can estimate
the disk parameters from the observed $\gamma$-ray luminosity
($L \sim 1.8 \times 10^{49}$~erg/s) and lifetime ($t \sim 1$~s). 
Eqs.~(\ref{LBZb}), (\ref{Mdota}) and (\ref{tdisk}) yield
\begin{equation}
\label{GW150914}
\frac{M_{\rm disk}}{M_{\rm BH}} \sim 1\times 10^{-5}, \ \ \frac{R_{\rm disk}}{M_{\rm BH}} \sim 20, 
\ \ {\dot M_{\rm BH}} \sim 0.9 \times 10^{-3}~M_{\odot}~{\rm s^{-1}} .
\end{equation}
Further simulations and future detections will be required to
assess this possibility.

\medskip 

{\it Acknowledgments}: It is a pleasure to thank E. Berger, C. Gammie, V. 
Paschalidis, M. Ruiz and L. Sun for useful discussions. This paper was 
supported in part by NSF Grants PHY-1300903 and PHY-1602536 
and NASA Grant NN13AH44G at the University of Illinois at
Urbana-Champaign. This work used the
Extreme Science and Engineering Discovery Environment (XSEDE), which
is supported by NSF Grant No. OCI-1053575. This research is part of
the Blue Waters sustained-petascale computing project, which is
supported by the National Science Foundation (Award No. OCI
07-25070) and the state of Illinois. Blue Waters is a joint effort of
the University of Illinois at Urbana-Champaign and its National Center
for Supercomputing Applications.

\bibliographystyle{apsrev4-1}  
\bibliography{paper}   

\end{document}